\documentclass[a4paper]{jpconf}
\usepackage{amssymb}
\usepackage{graphics}
\usepackage{amsthm}
\usepackage{latexsym}
\newcommand{\be}{\begin{eqnarray}}
\newcommand{\ee}{\end{eqnarray}}
\newcommand{\bdm}{\begin{displaymath}}
\newcommand{\edm}{\end{displaymath}}

\begin{document}
\title{\textbf{
Localizing gravity on thick branes: a solution for massive KK
modes of the Schr\"odinger equation}}
\author{\textbf{N Barbosa--Cendejas and A Herrera--Aguilar}}
\date{}
\address{Instituto de F\'{\i}sica y Matem\'{a}ticas, Universidad
Michoacana de San Nicol\'as de Hidalgo. \\
Edificio C--3, Ciudad Universitaria, C.P. 58040, Morelia,
Michoac\'{a}n, M\'{e}xico.} \ead{\textbf{nandinii@ifm.umich.mx,
herrera@ifm.umich.mx}} \vspace{1cm}
\begin{abstract}
We generate scalar thick brane configurations in a 5D Riemannian
space time which describes gravity coupled to a self--interacting
scalar field. We also show that 4D gravity can be localized on a
thick brane which does not necessarily respect $Z_2$--symmetry,
generalizing several previous models based on the Randall--Sundrum
system and avoiding the restriction to orbifold geometries as well
as the introduction of the branes in the action by hand. We begin
by obtaining a smooth brane configuration that preserves 4D
Poincar\'e invariance and violates reflection symmetry along the
fifth dimension. The extra dimension can have either compact or
extended topology, depending on the values of the parameters of
the solution. In the non--compact case, our field configuration
represents a thick brane with positive energy density centered at
$y=c_2$, whereas in the compact case we get pairs of thick branes.
We recast as well the wave equations of the transverse traceless
modes of the linear fluctuations of the classical solution into a
Schr\"odinger's equation form with a volcano potential of finite
bottom. We solve Schr\"odinger equation for the massless zero mode
$m^2=0$ and obtain a single bound wave function which represents a
stable 4D graviton and is free of tachyonic modes with $m^2<0$. We
also get a continuum spectrum of Kaluza--Klein (KK) states with
$m^2>0$ that are suppressed at $y=c_2$ and turn asymptotically
into plane waves. We found a particular case in which the
Schr\"odinger equation can be solved for all $m^2>0$, giving us
the opportunity of studying analytically the massive modes of the
spectrum of KK excitations, a rare fact when considering thick
brane configurations.
\end{abstract}

\section{Introduction, setup and solution}

The fact that we can live in a multidimensional space time with
infinite extra dimensions turns out to be compatible with present
time gravitational experiments. In such scenarios, gravity is
essentially 4--dimensional from the point of view of an observer
located at a 3--brane in which matter is confined, however, the
world can be higher dimensional with extended extra dimensions and
gravity can propagate in all of them. Multidimensional space times
with large extra dimensions turned out to be very useful when
addressing several problems of high energy physics like the
cosmological constant, dark matter and the mass hierarchy problem
\cite{rubakov}--\cite{lr} as well as the recent
non--supersymmetric string model realization of the Standard Model
at low energy with no extra massless matter fields
\cite{kokorelis}. The striking success of these higher dimensional
scenarios gave rise to several generalizations, including scalar
thick brane configurations \cite{dewolfe}--\cite{varios}. These
field configurations were generalized in the framework of Weyl
geometries for $Z_2$--symmetric manifolds in \cite{ariasetal}.
Moreover, localization of 4D gravity on thick branes that break
reflection symmetry was presented in \cite{bh2}--\cite{bh3}.

In this paper we begin by considering a 5--dimensional action
which describes gravity coupled to a bulk scalar field. In this
framework we obtain smooth brane solutions which respect 4D
Poincar\'e invariance, do not necessarily respect reflection
$Z_2$--symmetry and allow for both compact and non--compact
manifolds in the extra dimension. The structure of these brane
configurations depends on the topology of the extra dimension and
the value of the parameter $p(\xi)$. We further investigate linear
fluctuations of the metric around the classical background
solution. We show that 4D gravity can be described in our setup
since the analog quantum mechanical problem, a Schr\"odinger
equation with a volcano potential, for the transverse traceless
sector of the fluctuations of the metric yields a continuum and
gapless spectrum of KK states with a stable zero mode that
corresponds to the 4D graviton. Finally, we consider a particular
case in which the Schr\"odinger equation can be completely
integrated for all the massive modes of KK states and present
analytical expressions for them.

Let us start by considering the following 5--dimensional
Riemannian action
\begin{equation}
\label{action} S_5=\int_{M_5}\frac{d^5x\sqrt{|g|}}{16\pi
G_5}[R_5+3{\xi}(\nabla\omega)^2+6U(\omega)],
\end{equation}
where $\omega$ is a bulk scalar field, $\xi$ is an arbitrary
coupling parameter, and $U(\omega)$ is a self--interacting
potential for the scalar field.

We shall consider solutions which respect 4--dimensional
Poincar\'e invariance with the following line element
\begin{equation}
\label{conflinee} {ds}_5^2=e^{2\sigma(y)}\eta_{nm}dx^n
dx^m+e^{\omega(y)}dy^2,
\end{equation}
where $2\sigma(y)=2A(y)+\omega(y)$ depends only on the extra
coordinate $y$, and $m,n=0,1,2,3$. We shall call $e^{2A(y)}$ the
warp factor of the metric.

Further, by following \cite{ariasetal} we introduce the new
variables $X=\omega'$ and $Y=2A'$ and get the following pair of
coupled field equations from the action (\ref{action})
\begin{eqnarray}
\label{fielde} X'+2YX+\frac{3}{2}X^2=\frac{1}{\xi}\frac{d U}
{d\omega}e^{\omega},\nonumber\\
Y'+2Y^2+\frac{3}{2}XY=\left(-\frac{1}{\xi}\frac{d U}
{d\omega}+4U\right)e^{\omega}.
\end{eqnarray}
In general, it is not a trivial task to fully integrate these
field equations. However, it is straightforward to construct
several particular solutions to them after some simplifications
and, in special cases, these solutions lead to expressions of the
dynamical variables that can be treated analytically in closed
form, an important advantage from the physical point of view.

The system of equations (\ref{fielde}) can be easily solved if one
uses the condition $X=kY$, where $k$ is an arbitrary parameter which
is not allowed to adopt the value $k=-1$ because the system would
become incompatible. This condition restricts the self--interacting
potential to adopt the form $$U=\lambda
e^{\frac{4k\xi}{1+k}\omega}$$ and the field equations (\ref{fielde})
reduce to the following differential equation
\begin{equation}
\label{finale}
Y'+\frac{4+3k}{2}Y^2=\frac{4\lambda}{1+k}e^{(1+\frac{4k\xi}{1+k})\omega}.
\end{equation}

In \cite{ariasetal}--\cite{bh2} this equation was solved by choosing
the parameter $\xi=-(1+k)/(4k)$, while in \cite{bh3} another
solution was obtained by fixing the parameter $k=-4/3$. In this
paper we shall continue to consider the latter particular case.
Thus, after imposing the condition $k=-4/3$, the second term in the
left hand side of the equation (\ref{finale}) vanishes, yielding the
following differential equation
\begin{equation}
\omega''-16\lambda e^{p\omega}=0, \label{diffeqw}
\end{equation}
where $p=1+16\xi$.

One can solve this equation for $\omega$ and, by further integrating
the relation $2A'=-3\omega'/4$, one gets the solution
\begin{eqnarray}
\label{pairsolut} \omega=-\frac{2}{p}\ln\left[\frac{\sqrt{-8\lambda
p}}{c_1}\cosh\left(c_1(y-c_2)\right)\right], \nonumber \\
e^{2A}=\left[\frac{\sqrt{-8\lambda
p}}{c_1}\cosh\left(c_1(y-c_2)\right)\right]^{\frac{3}{2p}},
\end{eqnarray}
where $c_1$ and $c_2$ are integration constants.

For $p<0$, this solution represents a localized object which does
not necessarily preserve the reflection symmetry ($y\rightarrow
-y$) along the fifth dimension and breaks it through non--trivial
values of the shift parameter $c_2$. Thus, the 5--dimensional
space time is not restricted to be an orbifold geometry as in the
Randall--Sundrum (RS) case, allowing for a more general type of
manifolds. The topology of the extra coordinate can be compact or
extended depending on the signs of the constants $p$ and
$\lambda$, and the real or imaginary character of the parameter
$c_1$. This constant characterizes the width of the warp factor
$\Delta\sim1/c_1$.

Let us consider the cases of physical interest:

\noindent A) $\lambda>0$, $p<0$, $c_1>0$. In this case the domain
of the fifth coordinate is infinite $-\infty<y<\infty$ and we have
a non--compact manifold in the extra dimension. In this case the
warp factor is located around the point $y=c_2$ and represents a
smooth localized function of width $\Delta$ which reproduces the
metric of the RS model in the thin brane limit, namely, when
$c_1\rightarrow\infty$, $p\rightarrow-\infty$ keeping
$c_1/p=\beta$ finite. In \cite{ariasetal}--\cite{bh3} these smooth
configurations were physically interpreted as thick branes in the
framework of Weyl geometry.

\noindent B) $\lambda>0$, $p>0$, $c_1=iq_1$. In this case we have
a compact extra dimension with $-\pi\le q_1(y-c_2)\le\pi$. The
corresponding expressions for the warp factor and the scalar field
are the following
\begin{eqnarray}
e^{2A(y)}=\left[\frac{\sqrt{8\lambda
p}}{q_1}\cos\left(q_1(y-c_2)\right)\right]^{\frac{3}{2p}},\nonumber\\
\omega=-\frac{2}{p}\ln\left[\frac{\sqrt{8\lambda
p}}{q_1}\cos\left(q_1(y-c_2)\right)\right].
\end{eqnarray}

In \cite{bh3} it was shown that this solution describes, in fact,
a pair of thick branes located in two different disconnected
regions of the manifold due to the fact that the 5--dimensional
curvature scalar is singular at the points
$y=\pm\frac{\pi}{2q_1}+c_2$:
\begin{equation}
\label{R5c} R_5=\frac{14q_1^2}{p}\left[\frac{\sqrt{8\lambda p
}}{q_1}\cos(q_1(y-c_2))\right]^{-2/p}
\left[1+\frac{8p-27}{8p}\tan^{2}\left(q_1(y-c_2)\right)\right],
\end{equation}
where plausibly we have null scalar energy densities.

Other cases of physical interest are equivalent or contained in
cases A) and B).

\section{Fluctuations of the classical background}

Let us turn to study the metric fluctuations $h_{mn}$ of the
interval (\ref{conflinee}) given by the perturbed line element
\begin{equation}
\label{mfluct} ds_5^2=e^{2\sigma(y)}[\eta_{mn}+h_{mn}(x,y)]dx^m
dx^n+e^{\omega(y)}dy^2.
\end{equation}
In the general case, one cannot avoid considering fluctuations of
the scalar field when treating fluctuations of the background
metric since they are coupled, however, in \cite{dewolfe} it was
shown that the transverse traceless modes of the background
fluctuations decouple from the scalar sector. As a result, these
modes can be approached analytically in closed form.

In order to apply this method, we first perform the coordinate
transformation \be dz=e^{-A}dy.\label{coordtransf}\ee This change
of variable leads to a conformally flat metric and, hence, the
transverse traceless modes of the metric fluctuations $h_{mn}^T$
obey the following wave equation
\begin{equation}
\label{eqttm}
(\partial_z^2+3\sigma'\partial_z+\Box^{\eta})h_{mn}^T=0.
\end{equation}
It is easy to see that this equation supports a massless and
normalizable 4D graviton given by $h_{mn}^T=C_{mn}e^{ipx}$, where
$C_{mn}$ are constant parameters and $p^2=m^2=0$.

By following \cite{rs} we shall recast equation (\ref{eqttm}) into
a Schr\"{o}dinger's equation form. In order to accomplish this
aim, we adopt the following ansatz for the transverse traceless
modes of the metric fluctuations
$$h_{mn}^T=e^{ipx}e^{-3\sigma/2}\Psi_{mn}(z)$$ and get the
equation
\begin{equation}
\label{schrodinger} [\partial_z^2-V(z)+m^2]\Psi=0,
\end{equation}
where we have dropped the subscripts in $\Psi$, $m$ is the mass of
the KK excitation, and the quantum mechanical potential is
completely defined by the curvature of the manifold and reads
\begin{equation}
V(z)=\frac{3}{2}\partial_z^2\sigma+\frac{9}{4}(\partial_z\sigma)^2.
\end{equation}

In the non--compact case A) we have found two particular cases
($p=-1/4$ and $p=-3/4$) for which we can invert the coordinate
transformation (\ref{coordtransf}) and explicitly express the
variable $y$ in terms of the coordinate $z$. For the case $p=-3/4$
the functions $A(z)$ and $\sigma(z)$ adopt the form
$$A(z)=\ln\left[c_1/\sqrt{c_1^4(z-z_0)^2+6\lambda}\right]$$ and
$$\sigma(z)=\frac{7}{6}\ln\left\{c_1^2/\left[c_1^4(z-z_0)^2+6\lambda\right]\right\}.$$
The expression for $\sigma(z)$ yields the following explicit
formula for the analog quantum mechanical potential
\begin{equation}
\label{potentialncA} V(z)=
\frac{21c_1^4\left[3c_1^4(z-z_0)^2-4\lambda\right]}
{4\left[c_1^4(z-z_0)^2+6\lambda\right]^2}.
\end{equation}

In the framework of the Schr\"odinger equation, the spectrum of
eigenvalues $m^2$ parameterizes the spectrum of graviton masses
that a 4--dimensional observer located at $z_0$ sees. This
equation can be solved for the massless zero mode $m^2=0$ and the
only normalizable eigenfunction adopts the form
$$\Psi_0=q\left[c_1^4(z-z_0)^2+6\lambda\right]^{-7/4},$$ where $q$ is
a normalization constant. Thus, this function constitutes the
lowest energy eigenfunction of the Schr\"odinger equation
(\ref{schrodinger}) since it has no zeros. This fact allows for
the existence of a 4D graviton with no tachyonic instabilities
from transverse traceless modes with $m^2<0$. In addition to this
massless mode, there exists a tower of higher KK modes with a
continuum spectrum of positive masses $m^2>0$.

A similar situation takes place in the compact case B). Remarkably,
the coordinate transformation $dz=e^{-A}dy$ can be inverted for
$p=3/8$ yielding $$\cos[q_1(y-c_2)]=\pm
q_1/\sqrt{q_1^2+9\lambda^2(z-z_0)^2},$$ i.e., decompactifying the
fifth dimension and pushing to infinity the singularities. This
mathematical fact implies that we actually have two disconnected
regions in the manifold: the region $-\frac{\pi}{2}\le
q_1(y-c_2)\le\frac{\pi}{2}$ is separated from the region
$\frac{\pi}{2}\le q_1(y-c_2)\le\frac{3\pi}{2}$ (since we can shift
the domain of the compact dimension to $-\frac{\pi}{2}\le
q_1(y-c_2)\le\frac{3\pi}{2}$) by the physical singularities located
at $y=\pm\frac{\pi}{2q_1}+c_2$ (recall that the curvature scalar is
singular at these points). Each one of these regions leads to
$$A(z)=\ln\left\{3\lambda/[q_1^2+9\lambda^2(z-z_0)^2]\right\}$$ and
$$\sigma(z)=\frac{7}{3}\ln\left\{3\lambda/[q_1^2+9\lambda^2(z-z_0)^2]\right\}.$$
Thus, the analog quantum mechanical potential takes the form
\begin{equation}
\label{potentialncB}
V(z)=\frac{63\lambda^2\left[72\lambda^2(z-z_0)^2-q_1^2\right]}
{\left[q_1^2+9\lambda^2(z-z_0)^2\right]^2}
\end{equation}
and the wave function corresponding to the zero mode reads
$$\Psi_0=k\left[q_1^2+9\lambda^2(z-z_0)^2\right]^{-7/2},$$ where $k$
is a constant.

Both potentials (\ref{potentialncA}) and (\ref{potentialncB}) have
the volcano form: a well of finite bottom and positive barriers at
each side that vanish asymptotically. The corresponding wave
functions of the massless zero modes represent smooth lumps
localized around the point $z_0$. These facts imply that we have
only one gravitational bound state (the massless one) and a
continuous and gapless spectrum of massive KK states with $m^2>0$
in both cases A) and B).

Thus, we have obtained smooth brane generalizations of the RS
model with no reflection symmetry imposed in which the 4D
effective theory possesses an energy spectrum quite similar to the
spectrum of the thin wall case, in particular, 4D gravity turns
out to be localized at a certain value of the fifth dimension in
both cases A) and B).

Finally we would like to present the particular case $p=3/4$ in
which the coordinate transformation (\ref{coordtransf}) can be
inverted as well yielding the equality
$$
\cos\left[q_1(y-y_0)\right]={\rm
sech}\left[\sqrt{6\lambda}(z-z_0)\right].
$$
Here we have again two disconnected regions in the manifold due to
the presence of physical singularities at
$y=\pm\frac{\pi}{2q_1}+c_2$. In each one of these regions we
perform the change of variable (\ref{coordtransf}) and we observe
that this is equivalent to decompactifying one more time the fifth
dimension and pushing the singularities to spatial infinity. Thus
we get the following expressions for the warp functions
$$A(z)=\ln\left\{\frac{\sqrt{6\lambda}}{q_1}
{\rm sech}\left[\sqrt{6\lambda}(z-z_0)\right]\right\}$$ and
$$\sigma(z)=\frac{7}{3}\ln\left\{\frac{\sqrt{6\lambda}}{q_1}
{\rm sech}\left[\sqrt{6\lambda}(z-z_0)\right]\right\}.$$
Consequently, the analog quantum mechanical potential takes the
following form
\begin{equation}
\label{potentialpes3e4}
V(z)=\frac{21\lambda}{2}\left\{5\tanh^2\left[\sqrt{6\lambda}(z-z_0)\right]-2\right\}.
\end{equation}
It turns out that with this potential, the Schr\"odinger equation
can be integrated for both the massless and the massive modes of
the KK excitations. For the zero mode we get the following
normalizable wave function \be \Psi_0=C_0\ {\rm
sech}^{7/4}\left[\sqrt{6\lambda}(z-z_0)\right],\ee where $C_0$ is
a constant parameter.

On the other side, for the massive modes of the KK excitations we
get their expression in terms of the associated Legendre functions
of first and second kind: \be \Psi=C_1\
P_{7/2}^{\sqrt{147\lambda-2m^2}/\sqrt{12\lambda}}
\left[\tanh\left(\sqrt{6\lambda}(z-z_0)\right)\right]+\nonumber\ee
\be C_2\ Q_{7/2}^{\sqrt{147\lambda-2m^2}/\sqrt{12\lambda}}
\left[\tanh\left(\sqrt{6\lambda}(z-z_0)\right)\right],\ee where
$C_1$ and $C_2$ are integration constants.

This remarkable fact gives us the possibility of studying
analytically the massive modes of the spectrum of KK excitations,
a scarce phenomenon when considering smooth brane configurations.

\section{Concluding Remarks}

We considered the generation of scalar thick brane configurations
in a Riemannian manifold. We obtained a solution which preserves
4D Poincar\'e invariance and, in particular, represents a smooth
localized function characterized by the width parameter
$\Delta\sim 1/c_1$ and the constant $c_2$ which breaks the
$Z_2$--symmetry along the extra dimension; both of these
parameters are integration constants of the relevant field
equation (\ref{diffeqw}). Thus, our field configurations
correspond to smooth generalizations of the RS model which do not
restrict the 5--dimensional space time to be an orbifold geometry,
a fact that can be useful in approaching several issues like the
cosmological constant problem, black hole physics and holographic
ideas, where there is a relationship between the position in the
extra dimension and the mass scale.

We encountered two different cases regarding the topology of the
extra dimension: an extended and a compact one. In the
non--compact case A), the warp factor reproduces the metric of the
RS model in the thin brane limit, even if the matter content of
the theory does not correspond to the same brane configuration. In
the compact case B) the situation is different: we have several
pairs of thick brane configurations disconnected by physical
singularities. The structure of these branes depends on the value
of the parameter $p(\xi)$. In two special cases ($p=3/8$ and
$p=3/4$) we managed to invert the coordinate transformation
(\ref{coordtransf}) which makes the metric conformally flat,
decompactifies the fifth dimension and simultaneously pushes the
singularities of the manifold to infinity.

We wrote the wave equations of the transverse traceless modes of
the linear fluctuations of the classical background into a
Schr\"odinger's equation form for both cases A) and B). In
general, the analog quantum mechanical potential involved in it
represents a volcano potential with finite bottom: a negative well
located between two finite positive barriers that vanish when
$z\longrightarrow\pm\infty$. It turned out that for the massless
zero modes ($m^2=0$) the Schr\"odinger equation can be solved in
both cases. As a result of this fact, in each case we obtained an
analytic expression for the lowest energy eigenfunction of the
Schr\"odinger equation which represents a single bound state and
allows for the existence of a stable 4D graviton since there are
no tachyonic modes with $m^2<0$. Apart from these massless states,
we also got a continuum and gapless spectrum of massive KK modes
with positive $m^2>0$ that are suppressed at $y=c_2$ and turn
asymptotically into continuum plane waves in both cases A) and B),
as in \cite{lr}, \cite{dewolfe} and \cite{bh2}.

The shape of the analog quantum mechanical potential and the
localization of 4D gravity on smooth branes with a continuum and
gapless spectrum of massive KK modes are quite similar to those
obtained by \cite{lr}, \cite{dewolfe} and \cite{gremm}.

We finally found that for the particular case ($p=3/4$) the
Schr\"odinger equation can be integrated for both the massless and
the massive modes of the KK excitations. On one hand, for the zero
mode we get a normalizable wave function that represents again a
single bound state interpreted as a stable 4D graviton since it
has no tachyonic modes with negative squared mass $m^2<0$. On the
other hand, for the massive modes of the KK excitations we get
expressions in terms of associated Legendre functions of first and
second kind, a remarkable fact that open to us the possibility of
analytically investigating the massive spectrum of KK states, a
rather scarce phenomenon when studying thick brane configurations.
We are currently analyzing the normalizability of the massive
modes of the KK excitations (see \cite{norma1}--\cite{norma3}) and
hope to present our results in the near future.

\ack One of the authors (AHA) thanks the organizers of the NEB XII
Conference for providing a warm atmosphere plenty of interesting
and useful discussions in Nafplio. Both authors are really
grateful to R. Maartens, A. Merzon, U. Nucamendi, C. Schubert and
T. Zannias for fruitful discussions while this investigation was
carried out. This research was supported by grants CIC-UMSNH-4.16
and CONACYT-F42064.

\end{document}